\documentclass[10pt,twocolumn]{IEEEtran}
\usepackage{graphicx,amssymb,amsmath,amsbsy,bm,amsfonts}

\newtheorem{Theorem}{Theorem}

\newtheorem{Proposition}{Proposition}

\newtheorem{remark}{Remark}

\DeclareMathOperator{\tr}{tr}

\def\bfX{ {\boldsymbol X} }
\def\bfx{ {\boldsymbol x} }
\def\bfz{ {\boldsymbol z} }
\def\bfa{ {\boldsymbol a} }
\def\bfb{ {\boldsymbol b} }
\def\bfc{ {\boldsymbol c} }
\def\bfr{ {\boldsymbol r} }
\def\bfy{ {\boldsymbol y} }

\def\bftheta{ {\boldsymbol \theta} }
\def\bfepsilon{ {\boldsymbol \epsilon} }

\def\Pset{ {\mathsf{P}} }
\def\Pdist{ {{\cal{P}}_{Q, \mu}} }
\def\Dset{ {\mathsf{D}} }
\def\Yset{ {\mathsf{Y}} }
\def\Qset{ {\mathsf{Q}} }
\def\UnitH{ S^{n-1}}
\def\dif{ {\rm d} }

\def\Sprn{S(P) }
\def\Sprni{S(P_i) }
\def\Sprnj{S(P_j) }
\def\Sprnzero{{S(P_0) }}
\def\Sprnone{S(P_1) }

\def\EqualDist{\mathrel{\mathop=^{\rm d}}}
\def\EqualDef{\mathrel{\mathop=^{\rm def}}}

\begin{document}


\title{A Sampling Strategy for Projecting to Permutations in the Graph Matching Problem}

\author{R.~J.~Wolstenholme  and A.~T.~Walden, {\it Senior Member, IEEE} \thanks{
Rob Wolstenholme and Andrew Walden 
are both at the Department of Mathematics, Imperial College  London, 180 Queen's Gate,
London SW7 2BZ, UK.  
(e-mail: rjw08@imperial.ac.uk and a.walden@imperial.ac.uk)
\newline
}}

\maketitle

\begin{abstract} 
In the context of the graph matching problem
we propose a novel method for projecting a matrix $Q$, which may be a doubly stochastic matrix, to a permutation matrix $P.$ We observe that there is an intuitve mapping, depending on a given $Q,$ from the set of $n$-dimensional permutation matrices to sets of points in $\mathbb{R}^n$. The mapping has a number of geometrical properties that allow us to succesively sample points in $\mathbb{R}^n$ in a manner similar to simulated annealing, where our objective is to minimise the graph matching norm found using the permutation matrix corresponding to each of the points. Our sampling strategy is applied to the QAPLIB benchmark library and outperforms the PATH algorithm in two-thirds of cases. Instead of using linear assignment,
the incorporation of our sampling strategy as a projection step into algorithms such as PATH itself has the potential to achieve even better results.

\end{abstract}

\begin{keywords} Graph matching, permutation matrix, doubly stochastic matrix, sampling strategy.
\end{keywords}
\section{Introduction}
Graph matching  is important in many different areas of research \cite{Conte_etal04}. It is particularly well studied in the field of computer vision but has many other applications, ranging from circuit design to social network analysis. Exact graph matching consists of trying to find an exact isomorphism from one graph (or subgraph) to another. In inexact graph matching one aims to find the best permutation  of one of the graphs to make it as similar as possible to the other. This paper is concerned only with inexact graph matching and we refer to it henceforth simply as graph matching.

A paramount issue with graph matching is the fact that the number of fixed node arrangements for a graph is factorial in the dimension of the graph. The computation time for optimal accuracy algorithms  becomes computationally intractable as dimension increases \cite{BunkeAllermann83}. Instead many suboptimal methods have been developed to find a balance between speed and accuracy.

One approach uses spectral methods based on the graph Laplacian or adjacency matrix as eigenvalues and eigenvectors of both these matrices are invariant with respect to node permutation \cite{Knossow_etal09,Umeyama88}.

It is also possible to work directly with the adjacency matrices themselves, (e.g., \cite{AlmohamadDuffuaa93} and \cite{Zaslavskiy_etal09}). \cite{Zaslavskiy_etal09} introduces a convex-concave programming approach to give an approximate solution for labelled graph matching, a generalisation of graph matching. 
The paper identified that there are many cases where the `common approach' of 
\begin{itemize}
\item[(i)]{} relaxing the graph matching problem to find a solution $Q$ in a superset of the permutation matrices and 
\item[(ii)]{} projecting to the closest permutation matrix $P,$ (minimising Frobenius norm), 
\end{itemize}
does not find a satisfactory solution.

Instead the approach in \cite{Zaslavskiy_etal09} used a gradual updating of the initial solution $Q$ towards a solution in the set of permutation matrices, following a path calculated by the convex-concave programming approach. The procedure is known as the PATH algorithm. It combats the inefficiency of the previously mentioned approach by updating the relaxed solution $Q$. In this paper, by contrast, we  look at modifying the common approach by improving the second (projection) step.

In Barvinok \cite{Barvinok06} it was shown how an orthogonal matrix $Q$ could be approximated as a `non-commutative convex combination' of permutation matrices. In his proof he used the idea of randomised rounding to project $Q$ onto a permutation matrix $P$ so that $Q$ defines a distribution over the set of permutation matrices (along with the sampling distribution used for the randomised rounding). This  is also briefly mentioned in \cite{Fogel_etal13}.

In this paper we propose a sampling strategy for projecting matrices over ${\mathbb{R}}^{n \times n}$ 
to permutation matrices in the graph matching environment. In sections~\ref{sec:gmp}  and \ref{sec.currentMethod} we briefly discuss the graph matching problem, how it can be relaxed and the standard approach to finding an approximate solution.

In section~\ref{sec.ourMethod}
we show how the ideas in \cite{Barvinok06}   can be used to develop a graph matching strategy. We transform the problem of sampling in the space of permutation matrices to sampling  vectors $\bfx \in {\mathbb{R}}^{n}.$  Given a relaxed solution $Q,$ 
instead of solving $\arg \min_{P } || Q - P ||^2_{\rm F}$ 
we solve $\arg \min_{P} || Q \bfx - P \bfx ||^2_{\rm F}$
for a given $\bfx \in {\mathbb{R}}^{n}$. (Here $|| A ||_{\rm F}$ denotes the Frobenius norm
$
||A||_{\rm F}=[\tr\{A A^H\}]^{1/2}
,$ $\tr\{\cdot\}$ denotes trace,
$^H$ denotes complex-conjugate (Hermitian) transpose.)
We say that $\bfx$ is in the permutation set $S(P^*)$ if $P^*=\arg \min_{P} || Q \bfx - P \bfx ||^2_{\rm F}.$
We show that the solution to the `common approach' corresponds to minimizing the mean of the squared norm for our method, under uniform distributions on the unit hypersphere or unit cube. 

In section~\ref{sec.roundingGeometry} we investigate some geometrical properties of our proposal. We show there is a `degree of continuity,' i.e., given a point $\bfx\in \Sprn$, we can find other points close to it that are also in $\Sprn$. The boundaries of the $S(P)$ regions are illustrated. We also solve the `reversed' problem: if we have a permutation $P,$ can we find $\bfx$ such that $\bfx\in S(P).$ This is a very useful result for our final sampling algorithm. Section~\ref{subsec:varianceadapt} describes a procedure for adjusting the variance of our proposal distributions as time progresses, and this is built-in to the full sampling  strategy for projecting to permutations in the graph matching problem algorithm which is detailed
Section~\ref{subsec:strategy}. Finally, in section \ref{sec.Results}, we show how our scheme, SSQCV, performs on QAPLIB \cite{Burkhard_etal97} a popular library of benchmark cases to test against. We find that SSQCV outperforms PATH \cite{Zaslavskiy_etal09} in two-thirds of the experiments, and the latter is already known to outperform well-known competitors. We point out that
the incorporation of our sampling strategy as a projection step into algorithms such as PATH itself has the potential to achieve even better results.

\section{Graph Matching Problem}\label{sec:gmp}
\subsection{Definition} 
An $n$-dimensional graph is represented as $G=(V,E)$ where $V = \{1 \dots n \}$ is a set of vertices and $E \subset V \times V$ is a set of edges such that $(i,j) \in E$ if and only if there is a connection from vertex $i$ to vertex $j$. We consider simple undirected graphs such that there are no self loops and $(i,j) \in E \Longleftrightarrow (j,i) \in E$. The set of edges $E$ can be represented by an adjacency matrix $A \in {\mathbb{R}}^{n \times n}$ such that $A_{ij} = 1$ if $(i,j) \in E$ and $A_{ij} = 0$ if $(i,j) \notin E$.

Consider two graphs with $n \times n$ adjacency matrices $A$ and $B$ (weighted or un-weighted), then the graph matching problem is concerned with finding 
\begin{equation} \label{eqn.GMP}
	P^* = \arg \min_{P \in {\Pset}} ||A - P^T B P ||^2_{\rm F}
\end{equation}
where $\Pset = \{P \in {\mathbb{R}}^{n \times n} : P {\bf 1} = P^T {\bf 1} = {\bf 1} \mbox{ and } P_{ij} \in \{0, 1\} \}$ is the set of dimension-$n$ permutation matrices.

An exhaustive search over $\Pset$ can be used to solve (\ref{eqn.GMP}), but it has complexity $O(n!)$ and is computationally intractable even for moderately sized $n$.

\subsection{Relaxation}
An alternative approach is to relax the constraints in (\ref{eqn.GMP}) to first find
\begin{equation} \label{eqn.GMPrelax}
	Q^* = \arg \min_{Q \in \Yset} || A - Q^T B Q ||^2_{\rm F}
\end{equation}
for some set $\Yset \supset \Pset$.

We could use for example $\Yset = \Dset = \{Q \in {\mathbb{R}}^{n \times n} : Q {\bf 1} = Q^T {\bf 1} = {\bf 1} \mbox{ and } Q  \succeq 0\}$, the set of doubly stochastic matrices, i.e., all matrices with non-negative entries whose rows and columns sum to 1. In this case, the optimisation in (\ref{eqn.GMPrelax}) is convex and can be efficiently solved by the Frank-Wolfe algorithm \cite{FrankWolfe56}.

Alternatively we could use $\Yset = \Qset = \{Q \in {\mathbb{R}}^{n \times n} : QQ^T = I \}$, the set of orthogonal matrices. (\ref{eqn.GMPrelax}) can then be efficiently solved using the singular value decomposition (SVD). Note however there is an unidentifiablity issue in this case and we do not have a unique solution. In the best case, we have $2^n$ solutions but can have more if the eigenvalues of either of the adjacency matrices $A$ and $B$ are not distinct.

Now $\Dset$ is the convex hull of the permutation matrices and intuitively a matrix $P \in \Pset$ that is `close' to $Q^* \in \Dset$ is a good candidate for a solution to (\ref{eqn.GMP}). This suggests we can find a good approximation to a solution of (\ref{eqn.GMP}) via:

\begin{enumerate}
	\item Solve (\ref{eqn.GMPrelax}), (which can be done efficiently), to get matrix $Q^* \in \Dset$.
	\item Project/round the matrix $Q^*$ to the closest matrix $P \in \Pset$.
\end{enumerate}


\section{Matrix Rounding --- Current Method} \label{sec.currentMethod}
After finding a suitable $Q$ for a relaxed version of (\ref{eqn.GMP}), i.e., $Q$ solving (\ref{eqn.GMPrelax}), the most common method to project  to a permutation is through the intuitive optimisation (see e.g., \cite{Zaslavskiy_etal09} ),
\begin{equation} \label{eqn.matrixRoundCurrent}
	\arg \min_{P \in \Pset} || Q - P ||^2_{\rm F} = \arg \max_{P \in \Pset} \,\tr\{Q^{T} P\}  
\end{equation}
which can be solved by the Hungarian algorithm in $O(n^3)$ time as $\max_{P \in \Pset} \tr (Q^{T} P)$ is simply a linear assignment problem \cite{Burkhard_etal09}.

A serious issue with the use of (\ref{eqn.matrixRoundCurrent}) is that it only delivers one candidate solution to (\ref{eqn.GMP}) and if it is not a good solution it is unclear how to continue. In \cite{Zaslavskiy_etal09} an incremental improvement to their estimate is performed by subsequent concave and convex relaxations, while still using  (\ref{eqn.matrixRoundCurrent}). We instead propose an adjustment to the projection step in itself as an alternative to (\ref{eqn.matrixRoundCurrent}).

\section{Matrix Rounding --- Proposed Method} \label{sec.ourMethod}
An important part of our overall algorithm, that we use to replace (\ref{eqn.matrixRoundCurrent}), involves solving
\begin{equation} \label{eqn.matrixRoundOurs}
	\arg \min_{P \in \Pset} || Q \bfx - P \bfx ||^2_{\rm F}
\end{equation}
for a given $\bfx \in {\mathbb{R}}^{n}$.
\subsection{Barvinok's Method}
The idea is inspired by Barvinok in \cite{Barvinok06}: to round an orthogonal matrix $Q$ to a permutation matrix $P$, consider its action on $\bfx \in {\mathbb{R}}^n$ sampled from a Gaussian distribution.
Consider sample $\bfx$ and ordering vector $\bfr (\bfx)$ such that $\bfr (\bfx) _i = j$ where $x_i$ is the $j$th smallest value of $\bfx$. For example:
$ 
\bfx = 
\begin{bmatrix}
         3.1, 7.3, 2.4, 8.7 
\end{bmatrix}^T
\Rightarrow
\bfr (\bfx) = 
\begin{bmatrix}
         2, 3,1,4 
\end{bmatrix}^T.$
Then Barvinok argues the permutation $P$ such that 
\begin{equation} \label{eqn.Barvinok}
	P \bfr(\bfx) = \bfr(Q \bfx)
\end{equation}
is `close' to $Q$ with respect to $\bfx$ as they both transform $\bfx$ in similar ways. $P$ represents a `rounding' of $Q$. 
Note also 
\begin{equation}\label{eq:roundstoself}
P \bfr(\bfx) = \bfr(P \bfx)
\end{equation} so if $Q$ is a permutation matrix, it is always rounded to itself.

This therefore provides a way to project/round an orthogonal matrix $Q$ to a distribution of permutation matrices. The distribution can be sampled from by drawing a Gaussian vector $\bfx \in {\mathbb{R}}^n$ and solving (\ref{eqn.Barvinok}).

We observe that for Barvinok's approach (i)
$Q$ need not be orthogonal, and (ii)
the distribution from which $\bfx$ is sampled need not be Gaussian.
Furthermore, we note the following important result:

\begin{Theorem} \label{thm.minSort}
	Given $\bfx \in {\mathbb{R}}^n$ and $Q \in {\mathbb{R}}^{n \times n}$,
any permutation matrix $P$ solving (\ref{eqn.Barvinok})
	is also a solution to (\ref{eqn.matrixRoundOurs}).
\end{Theorem}
\begin{IEEEproof}
This is given in Appendix~\ref{proof:ProofminSort}.
\end{IEEEproof}

\subsection{Solution and Effects of Scale of $\bfx$}
\begin{Proposition}
The solution of (\ref{eqn.matrixRoundOurs}) is invariant to the norm of $\bfx,$ i.e., if $P(Q, \bfx)$ is the solution and we write $\bfx = (r, \bftheta)$ in polar coordinates then we can equally write $P(Q, \bftheta)$ as the solution.
\end{Proposition}
\begin{IEEEproof}
Consider $\bfx_1, \bfx_2 \in {\mathbb{R}}^{n}$ such that $\bfx_1 = (r_1, \bftheta)$ and $\bfx_2 = (r_2, \bftheta)$. Then as both the sorting of a vector $\bfx \in {\mathbb{R}}^n$ is unchanged by multiplication by some constant $k>0$ and also $Q(k \bfx) = k Q \bfx$ is also unchanged with respect to sorting, if $P$ solves (\ref{eqn.matrixRoundOurs}) for a given $Q$ and $\bfx_1$, it also solves it for $({r_2}/{r_1})\bfx_1 = \bfx_2,$
a rescaled version of $\bfx_1.$
\end{IEEEproof}
\begin{remark}
This means we can always sample $\bfx\in {\mathbb{R}}^{n}$ on the unit hypersphere.
\end{remark}

\subsection{Permutation Distribution}
We can now define a probability distribution over our permutation matrices via a random variable $\Pdist$ such that for possible sample outcomes $\bfX=\bfx\in {\mathbb{R}}^{n },$
$$\Pr(\Pdist = P) {\,\,\displaystyle{\EqualDef}\,\,} \Pr (\bfx \in S_Q(P)),
$$
where $\mu$ is the cumulative distribution function for the random variable 
$\bfX$ from which $\bfx$ is drawn, $Q \in {\mathbb{R}}^{n \times n}$ and 
\begin{equation} \label{eqn.permPartition}
	S_Q(P^*) = \{ \bfx_0 \in {\mathbb{R}}^{n }: P^* = \arg \min_{P \in \Pset} || Q \bfx_0 - P \bfx_0||^2_{\rm F} \}.
\end{equation}
Any sets of $\bfx$ in multiple $S_Q({P})$'s are sets of measure zero as we shall show later. 

We will drop the $Q$ from the $S_Q$ notation from now on unless it is unclear to which $Q$ matrix we are referring.

So the distribution of permutation matrices from which we want to sample candidates to solve (\ref{eqn.GMP}) is affected by both $Q \in {\mathbb{R}}^{n \times n}$ and the distribution $\mu$ from which we sample $\bfx.$ 

\subsection{Uniform Distribution and Current Method of \S \ref{sec.currentMethod}}
For $\bfx \in {\mathbb{R}}^n$ we now show that the solution to the `current method' of section~\ref{sec.currentMethod} corresponds to minimizing the mean squared norm,
$E( || Q \bfX - P \bfX ||^2_{\rm F}),$ when $\bfX$ is uniformly distributed on the unit hypersphere or in the unit hypercube. 
\begin{Proposition}\label{prop:meanequiv}
	For $\bfX$ uniformly  distributed on the unit hypersphere $\UnitH {\,\displaystyle{\EqualDef}\,} \{\bfx\in {\mathbb{R}}^n: ||\bfx||=1\}$,
	\begin{equation} \label{eqn.unifSphere}
		\arg \min_{P \in \Pset} ||Q - P||^2_{\rm F} = \arg \min_{P \in \Pset} E( || Q \bfX - P \bfX ||^2_{\rm F}).
	\end{equation}
\end{Proposition}
\begin{IEEEproof}
This is given in Appendix~\ref{proof:ProofExpectation}.
\end{IEEEproof}
\medskip
\begin{Proposition}\label{prop:hypercube}
	For $\bfx \in {\mathbb{R}}^n$ with $\bfX$ uniformly distributed in the unit hypercube $H=[0,1]^n$ for which $0 \leq x_i \leq 1$,
	\begin{equation}
		\arg \min_{P \in \Pset} ||Q - P||^2_{\rm F} = \arg \min_{P \in \Pset} E( || Q \bfX - P \bfX ||^2_{\rm F}).	
	\end{equation}
\end{Proposition}
\begin{IEEEproof}
This is given in Appendix~\ref{proof:ProofHypercube}.
\end{IEEEproof}

\begin{figure}[t!]
\begin{center}
\hspace{-0.1in}
\includegraphics[scale=0.58]{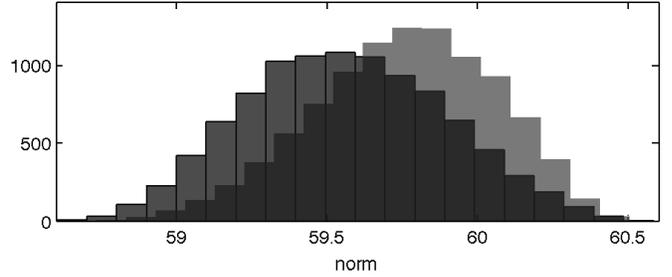}
\end{center}
\caption{Values of $||A - P_i^T B P_i ||_{\rm F}$ for uniform sampling of permutations  (light grey) versus sampling in ${\mathbb{R}}
^n$ and corresponding permutation sets (dark grey); the black is the overlap region.
}
\label{fig:UnifVsRnL0}
\end{figure}

\subsection{Illustrative Example}
Here we compare values of  $||A - P_i^T B P_i ||_{\rm F}, i=1,\ldots,m,$ obtained when 
\begin{itemize}
\item we sample $m$  independent outcomes of  $\bfx\in {\mathbb{R}}^n,$ the components $x_i$ being independent and  
having the standard normal distribution, and for each $\bfx$ solve (\ref{eqn.matrixRoundOurs}) to obtain the $P_i$'s, and
\item
we 
randomly sample $m$ $P_i$'s uniformly on $\Pset$ using the $\text{MATLAB}^\circledR$ function `randperm,' followed by conversion of the permutation sequence to a matrix. 
\end{itemize}
We would like to see that our sampling approach based on some permutation approximation $Q \in {\mathbb{R}}^{n \times n}$  dividing ${\mathbb{R}}^{n}$ 
into permutation sets $S(P) $  is better than simply randomly sampling permutations uniformly.

We let $A$ be a $15$-dimensional random symmetric matrix, $P$ be a $15$-dimensional random permutation matrix and $B = (P+\gamma Z) A (P+\gamma Z)^T$ where $Z_{ij} \displaystyle{\EqualDist{\,}} {\cal N}(0,1) $ (the normal distribution with mean zero and variance one) and $\gamma=0.1.$ We then find $Q$ solving (\ref{eqn.GMPrelax}) over $\Dset$. 

For this example,
the histograms of Fig.~\ref{fig:UnifVsRnL0} show that the distribution of the error norm for permutations sampled by solving (\ref{eqn.matrixRoundOurs}) is shifted to the left compared with sampling permutations randomly.

\section{Partitioning of ${\mathbb{R}}^n$}\label{sec.roundingGeometry}
\subsection{`Continuity'}
Here we consider the `continuity' (though not necessarily connectedness) of the sets $\Sprn$.
\begin{Proposition} \label{prop.closedSet}
For $\bfx \in S(P) $ such that $x_i \neq x_j$ and $(Q \bfx)_i \neq (Q \bfx)_j$ for $i \neq j$, we can find  $\epsilon >0$ such that ${\mathbb{B}}_\epsilon(\bfx) \subset \Sprn$, 
whenever ${\mathbb{B}}_\epsilon(\bfx) = \{ \bfy \in {\mathbb{R}}^n: ||\bfy - \bfx||^2_{\rm F} < \epsilon \}$.
\end{Proposition}
\begin{IEEEproof}
This is given in Appendix~\ref{app:closedSet}.
\end{IEEEproof}

This means that given a point $\bfx$ inside $\Sprn$ (not on its boundary), we can find other points close to it that are also in $\Sprn$. We can also see that the boundaries of $\Sprn$ occur at points where $x_i = x_j$ or $(Q \bfx)_i = (Q \bfx)_j$ for $i \neq j$. We may of course have multiple elements becoming equal at the same time, e.g. $\bfx = 0$ where $\bfx \in \Sprn$ for all permutation matrices.

\begin{Proposition}
If we are sampling from a purely continuous distribution with $\bfx \in {\mathbb{R}}^n$ defined by random variable $\bfX$, then
$ {\rm Pr}((\bfX = \bfx) \cap (x_i = x_j: i \neq j)) = 0 $
and
$ {\rm Pr}((\bfX = \bfx) \cap ((Q \bfx)_i = (Q \bfx)_j: i \neq j, Q \in \Dset \cup \Qset)) = 0. $
\end{Proposition}
\begin{IEEEproof}
	In both cases the sets are of measure zero in ${\mathbb{R}}^n$ and hence correspond to zero probability.
\end{IEEEproof}
So when we are sampling, we do not have to worry about hitting a point of discontinuity between two sets $\Sprni$ and $\Sprnj$ for $i \neq j$.
\begin{remark}
When we do change from $\Sprnzero$ to $\Sprnone$ in moving a `small' distance from point $\bfx^0$ to $\bfx^1$, the difference between $P_0$ and $P_1$ is normally small, (but not always, considering the case $\bfx = {\bf 0}$),  and occurs because for some $i$ and $j$, $x_i^0 < x_j^0$ but now $x_i^1 > x_j^1$. If $i$ and $j$ are the only entries that have flipped, (which happens in the majority of cases), then this change corresponds to a simple flip in the entries of $P_0$ and $P_1,$ e.g., if $P_0$ sent $1 \rightarrow 2$ and $3 \rightarrow 4$, $P_1$ may now send $1 \rightarrow 4$ and $3 \rightarrow 2$.  This means $||P_0 - P_1||^2_{\rm F} = 4,$ (the minimum such value between two different permutation matrices).
\end{remark}
\begin{remark}
There  exist boundary hyperplanes  between $\Sprnzero$ and $\Sprnone$ where $x_{i_1} = x_{i_2} = \dots = x_{i_k}$ for large $k$, so that crossing such hyperplanes results in $P_0$ and $P_1$ being quite dissimilar.
\end{remark}
This form of continuity outlined implies we can use a search algorithm based on closeness in ${\mathbb{R}}^n,$ which we detail in the next section.

\subsection{3D Sphere Visualisation}
In order to gain further insight into the partitioning of ${\mathbb{R}}^n$ into permutation sets, we will illustrate the case when $n=3$. 

The boundaries of the permutation sets are defined by the lines on the unit hypersphere given by $x_i = x_j$ and $(Q \bfx)_i = (Q \bfx)_j$ for a partition-defining matrix $Q \in {\mathbb{R}}^{n \times n}$.

We note that in the case $Q {\bf 1} = c {\bf 1}$ for $c \in \mathbb{R},$ e.g., $Q$ a doubly stochastic matrix, all the boundaries of the permutation sets intersect at the point $\bfa = 3^{-1/2} {\bf 1}$. The size of each permutation set can then be calculated using the angles between these boundary lines at the point $\bfa$. The ratio of the area of a permutation set with boundary angle $\theta$ to the whole unit sphere area is then ${\theta}/{2 \pi}$ and if we sample a point on the unit sphere, this is the probability of it being from inside the given permutation set. The angles $\theta$ can be readily found by considering the normal vectors to the planes that form the boundaries of the permutation sets in $\mathbb{R}^3,$ i.e., $x_i = x_j$ and $(Q \bfx)_i = (Q \bfx_j)$.

%
%
\begin{figure}[t!]
\begin{center}
\hspace{-0.2in}
\begin{math}
\begin{matrix}
\includegraphics[scale=0.9]{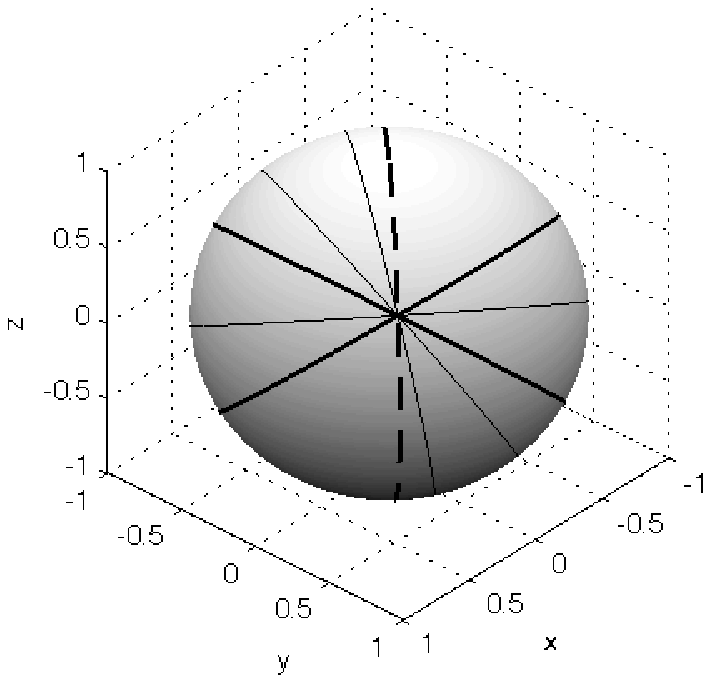}\\
\includegraphics[scale=0.9]{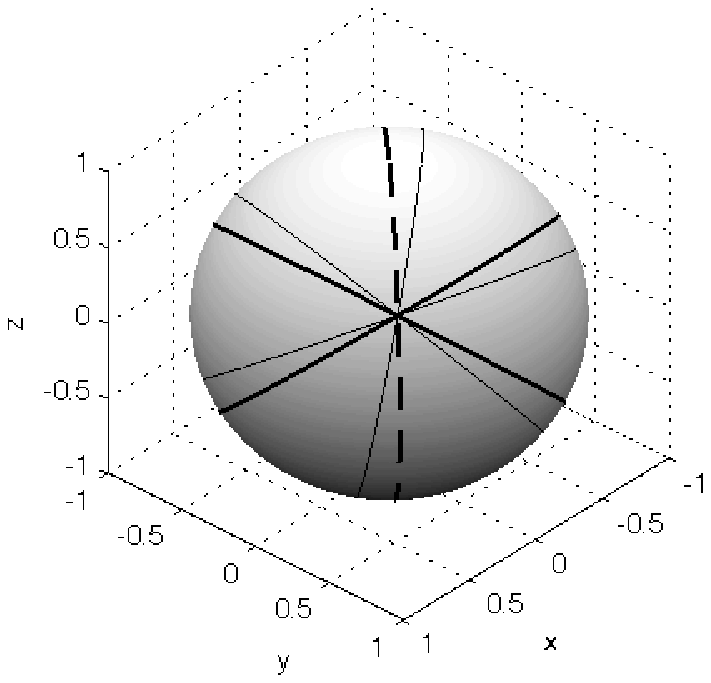}
\end{matrix}
\end{math}

\end{center}
\caption{3D sphere showing the partition boundaries for  a doubly stochastic matrix $Q$ (top figure), and for an orthogonal matrix (bottom figure). Here $x=x_1,y=x_2,z=x_3.$ See text for further details.
}
\label{fig.sphereDS}
\end{figure}

Our figures show the partition boundaries for random symmetric 3D matrices $A$ and $B$. These boundaries are the lines $(Q \bfx)_i = (Q \bfx)_j$ when $Q=I_n$ (heavy lines) and

\begin{itemize}
  \item Fig.~\ref{fig.sphereDS}(top): $Q$ is the best doubly stochastic matrix solving (\ref{eqn.GMPrelax}) (thin lines).
  \item Fig.~\ref{fig.sphereDS}(bottom): $Q$ is an orthogonal matrix with an eigenvector ${\bf 1}$ (thin lines).
\end{itemize}
The dashed line is for $x_1=x_2.$ 
We make the following observations.
When $Q=I_n$ the lines have a constant angle between them and pass through the point $\bfa = 3^{-1/2} {\bf 1}$.
When $Q$ is the best doubly stochastic matrix, the lines no longer have a constant angle between them but still pass through the point $\bfa$ because $Q \bfa = \bfa$. When $Q$ is an orthogonal matrix with an eigenvector ${\bf 1}$, the lines have a constant angle between them and pass through the point $\bfa$. We can think of this case as a rotating of the lines from the $Q=I_n$ case.

\subsection{A Reversal: Permutations to Points}
Our algorithm will make use of the following step: 
if we have a permutation $P,$ can we find an $\bfx$ that rounds to $P$ using (\ref{eqn.matrixRoundOurs})? The answer is yes, from the following result.

\begin{Theorem} \label{thm.reversePerm}
Consider $Q \in {\mathbb{R}}^{n \times n}$ and let $\bfa = n^{-1/2} \boldsymbol{1}.$ If
$ \bfb = Q^{-1} \bfa $
is such that $b_i = b_j \Rightarrow i = j$, then for any $P^* \in \Pset$, we can find $\bfx \in {\mathbb{R}}^n$ such that
$$ P^* = \arg \min_{P \in \mathbb{P}} || Q \bfx - P \bfx ||_{\rm F}^2 $$
and it is given by
$ \bfx = \bfb + Q^{-1} P^* P_b^T \bfepsilon, $
where $P_b$ orders $\bfb$ in ascending order (i.e. $\bfr((P_b \bfb))_i = i$) and $\bfepsilon = \delta[1, 2, \dots, n]^T$ for some $\delta >0$.
\end{Theorem}
\begin{IEEEproof}
This is given in Appendix~\ref{app:reversePerm}.
\end{IEEEproof}
We make the following observations.
\begin{itemize}
\item
 We can use Theorem~\ref{thm.reversePerm} to find initial points for our algorithm. Suppose $P_0 = \arg \min_{P \in \mathbb{P}} ||Q-P||^2_{\rm F}.$ Then replace $P^*$ by $P_0$ in Theorem~~\ref{thm.reversePerm} to find an $\bfx$ such that
\begin{equation}\label{eq:plugin}
 P_0=\arg \min_{P \in \mathbb{P}} || Q \bfx - P \bfx||^2_{\rm F}. 
\end{equation}
\item
 In general we want to choose $\delta>0$ to be as large as possible while still keeping $\bfr(\bfx) = \bfr(\bfb)$. This is because it moves $Q \bfx$ away from $\bfa$ which is a large point of discontinuity in our partitioned ${\mathbb{R}}^n$. The closer we are to it, the less positive  the effect on our algorithm will be from picking a suitable initial point.
\item
If $Q$ is a doubly stochastic matrix, it does not satisfy the conditions in Theorem~\ref{thm.reversePerm} because 
\begin{equation}\label{eq:problem}
Q \bfa = \bfa\Rightarrow\bfa=\bfb\,\,\,\text{and} \,\, \therefore\,\,\, b_i=b_j \not\Rightarrow i=j.
\end{equation}
\item If $b_i = b_j$ for some $i \neq j$, $\bfr ( \bfb )$ has no distinct ordering and multiple permutations sort $\bfb$ into ascending order.
\end{itemize}

\subsection{Doubly Stochastic Matrices}
It is possible to slightly perturb doubly stochastic matrix $Q\in \Dset$ by working instead with matrix 
\begin{equation}\label{eq:Qdash}
Q^{'} = Q + \lambda U,\qquad \lambda \in \mathbb{R},U \in {\mathbb{R}}^{n \times n},
\end{equation}
where $U$ is a matrix such that $U_{ij} \displaystyle{\EqualDist{\,}} {\rm Unif}[0,1]$. $Q^{'}$ avoids (\ref{eq:problem}) and so by Theorem~\ref{thm.reversePerm}, with probability 1, every permutation set has non-zero measure   --- some matching $\bfx$'s are guaranteed --- and it is therefore possible to sample all permutations in ${\mathbb{R}}^n$.

\begin{figure}[t!]
\begin{center}
\hspace{-0.2in}
\begin{math}
\begin{matrix}
\includegraphics[scale=0.9]{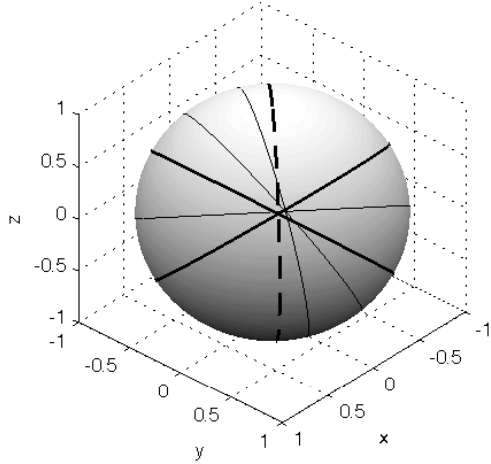}
\end{matrix}
\end{math}
\end{center}
\caption{3D sphere showing the partition boundaries for a perturbed doubly stochastic matrix $(\lambda=0.05).$
}
\label{fig.sphereDSPert}
\end{figure}

Fig.~\ref{fig.sphereDSPert} is of the same form as Fig.~\ref{fig.sphereDS} but now using $Q'.$ The lines no longer have a constant angle between them nor pass through the point $\bfa$.

\begin{figure}[t!]
\begin{center}
\includegraphics[scale=0.58]{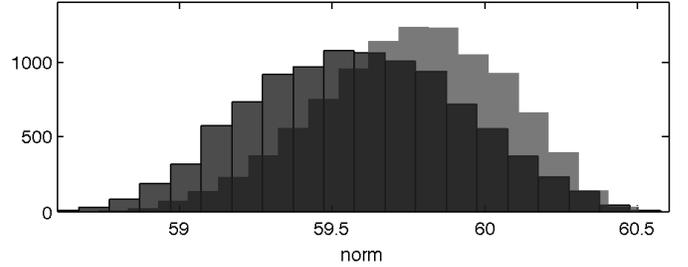}
\end{center}
\caption{Values of $||A - P_i^T B P_i ||_{\rm F}$ for uniform sampling of permutations  (light grey) versus sampling in ${\mathbb{R}}
^n$ and corresponding permutation sets (dark grey); the black is  the overlap region.
Here $Q\leftarrow Q+\lambda U.$ 
}
\label{fig:UnifVsRnL01}
\end{figure}

Fig.~\ref{fig:UnifVsRnL01} repeats Fig.~\ref{fig:UnifVsRnL0} only this time perturbing $Q$ to $Q'$ 
using $\lambda=0.1.$ 
While still
outperforming uniform sampling, the advantage has been slightly reduced.

\section{Variance Adaptation}\label{subsec:varianceadapt}
Before outlining our full sampling strategy, we  discuss an important component, namely `variance adaptation.'

Let $t\in {\mathbb{N}}$ denote a time step. 
In our sampling strategy, given $\bfx_t$, we obtain sample $\bfx_{t+1}$  from our proposal distribution $ {\cal N}_n(\bfx_t, \sigma^2_t I_n),$ 
the $n$-dimensional normal distribution with mean $\bfx_t$ and covariance matrix  $\sigma^2_t I_n.$

We then project the sample $\bfx_{t+1}$  onto the unit hypersphere using $\bfx_{t+1} \leftarrow \bfx_{t+1}/||\bfx_{t+1}||$. 
The only value we have control over is $\sigma^2_t$ and we investigate how best to choose this value so that the resultant samples have certain desirable properties.

Given a $Q \in {\mathbb{R}}^{n \times n}$, both $\bfx_t$ and $\bfx_{t+1}$ have associated permutation matrices $P(\bfx_t)$ and $P(\bfx_{t+1})$ respectively, from solving (\ref{eqn.matrixRoundOurs}).
Define 
\def\Pdiff{\Delta}
\def\modelP{\bar{\Delta}}
\def\tmodelP{\tilde{\Delta}}
\begin{equation}\label{eq:Pdifforig}
 \Pdiff_t = ||P(\bfx_t) - P(\bfx_{t+1})||_{\rm F} .
\end{equation}
Then $\sigma^2_t$ should be chosen so that
$\Pdiff_t$ gradually decays toward $0$ with increasing iteration step $t.$

Assume that 
\begin{equation} \label{eqn.PdiffEst}
	\Pdiff_t = \modelP(\sigma^2_t) + \epsilon_t,
\end{equation} 
where $\modelP(\sigma^2_t)$ is a model for $\Pdiff_t$ and $\{\epsilon_t\}$ is zero mean noise, so that $E(\Pdiff_t) = \modelP(\sigma^2_t)$. 

Suppose we can supply a target function $f_t$ 
 for $\Pdiff_t$ to follow. 
Then, at time $t$, given $f_t$ and $\modelP(\sigma^2_t)$ we can choose the value for $\sigma^2_t$ by calculating

\begin{equation} \label{eqn.chooseSigma}
	\sigma^2_t = \arg\min_{\sigma^2_{0t} > 0} ||f_t -\modelP(\sigma^2_{0t})||_{\rm F},
\end{equation}
i.e., the variance that minimises the distance between the target function and $\modelP$.

Make the substitution $y_t = \log(\sigma^2_t)$ since $\sigma^2_t > 0 $. 
Let $\Pdiff_{\rm max}$ be the maximum value of $\Pdiff_t.$ We then learn an estimator $\tilde \Pdiff(y_t)$ of $\modelP(y_t)/\Pdiff_{\rm max},$ using the observations $\Pdiff_t/\Pdiff_{\rm max}$ and $y_t.$ The function $\tmodelP(y_t)$ is taken to be
the logistic curve and is estimated via regression.

Consider approximating $\Pdiff_t$ for very large variance values. This can be done by randomly sampling $M$ points $\bfz_1, \dots \bfz_{M}$ on the unit hypersphere and finding 
\begin{equation} \label{eqn.Pmax}
	{\Delta}_{\rm max} \EqualDef \frac{1}{M} \sum_{i=1}^{M} ||P(\bfx_0) - P(\bfz_i)||_{\rm F},
\end{equation}
where $\bfx_0$ is the initial original unit hypersphere sample value. 
From (\ref{eq:Pdifforig}) and (\ref{eqn.PdiffEst}), 
$	{\Delta}_{\rm max} $ is an estimate for the largest value of $\modelP(\sigma^2_t).$ 

\subsection{Generating the Pre-samples}\label{subsec:presamples}
We need to generate a set of suitable $y_i$ values, which we label
 $y_{-L}, \ldots, y_{-1},$ called pre-samples.
In order to generate pre-samples, we want to sample a number of $y_i$ such that $\Pdiff_i/\Pdiff_{\rm max}\in[\epsilon, 1 - \epsilon]$ for some small chosen $\epsilon >0$. (This is to avoid a regression where all  $\Pdiff_i/\Pdiff_{\rm max}$ are either in $[1 - \epsilon,1]$ or $[0, \epsilon]$, in which case we are lacking information for accurately learning the logistic relationship.)

To do this we sample from a normal distribution with mean 0 and variance 1 to get our first point $y_a$ such that the corresponding $\Pdiff_a/\Pdiff_{\rm max}\notin [\epsilon, 1 - \epsilon].$  If it is in this interval, we increase the variance of our sampling distribution and keep trying until we get a suitable $y_a$. Once we have $y_a$ we then aim to find $y_b$ in a similar manner such that $[\min(\Pdiff_a,\Pdiff_b), \max(\Pdiff_a, \Pdiff_b)]/\Pdiff_{\rm max} \supset [\epsilon, 1 - \epsilon]$. We then reorder $y_a$ and $y_b$ such that for simplicity $y_a < y_b$.
We now sample the remaining $y_i$ uniformly on $[y_a, y_b]$. If $\Pdiff_i/\Pdiff_{\rm max} \notin [\epsilon, 1 - \epsilon]$, we update our sampling interval: if $\Pdiff_i/\Pdiff_{\rm max}  < \epsilon$, set $y_a = y_i$ else if $\Pdiff_i/\Pdiff_{\rm max}  > 1 - \epsilon$ set $y_b = y_i$.

After generating $L$ such $y$'s we call them $y_{-L}, \ldots, y_{-1}.$

\subsection{The Learning Step}
 We start with the $L$ `pre-samples' 
\begin{equation}\label{eq:ypresamps}
y_{-L}, \ldots, y_{-1},
\end{equation} 
and then
compute the corresponding $\bfx_{-i} \displaystyle{\EqualDist{\,}} {\cal N}_n(\bfx_0, {\rm e}^{y_{-i}}I_n),$ and the associated 
\begin{equation}\label{eq:minusis}
\Pdiff_{-i} = ||P(\bfx_0) - P(\bfx_{-i})||_{\rm F}.
\end{equation}
 We then learn $\tilde \Pdiff(y_t)$ via the inputs $y_{-L}, \ldots, y_{-1}$
and  $\Pdiff_{-L},\ldots,\Pdiff_{-1}.$

  We can also include a further parameter $T$ such that when $t = 0 \mod T$, we re-learn $\tmodelP$ given our observations up to that point. E.g., when $t=T$ we make use of inputs $y_{-L}, \ldots, y_{-1}, y_{1},\ldots,y_{T}$
and  $\Pdiff_{-L},\ldots,\Pdiff_{-1},\Pdiff_1,\ldots,\Pdiff_T$ to re-learn $\tmodelP.$

\begin{figure}[t!]
\begin{center}
\includegraphics[scale=1]{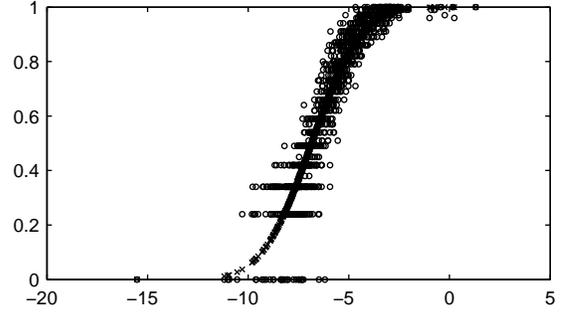}
\end{center}
\caption{Illustration of learning $\tmodelP(y_t).$ The  circles show
${\Pdiff_t}/{{\Delta}_{\rm max}}$ 
 (vertical) against $y_t$ (horizontal). The  crosses give the fitted logistic curve for  $\tmodelP(y_t).$
}
\label{fig.LogRegFit}
\end{figure}

Fig.~\ref{fig.LogRegFit} gives an example of the learning of $\tmodelP,$
and suggests a logistic model works appropriately.

\subsection{Target Function}
\begin{figure}[t!]
\begin{center}
\includegraphics[scale=1]{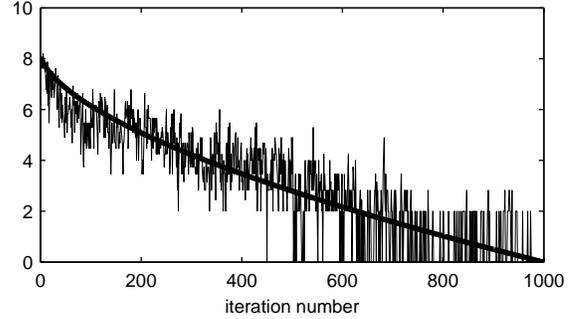}
\end{center}
\caption{$\Pdiff_t$ (hashy line) versus $f_t$ (thick curve), against $t.$
}
\label{fig.VarAdjustDecay}
\end{figure}

Fig.~\ref{fig.VarAdjustDecay} shows that we were able to choose $\sigma^2_t$ so that $\Pdiff_t$ does a good job of tracking the target curve $f_t$ defined here as $f_t=\Pdiff_{\rm max} [1-(t/1000)^{0.6}].$ 
This figure uses the re-learning step with $T=100.$ Particularly initially, $\Pdiff_t$ drifts below $f_t$ but this is corrected.


\section{Sampling Strategy with Variance Adaptation}\label{subsec:strategy}
The algorithm has characteristics in common with simulated annealing. In its purist form, we gradually sample proposal points closer and closer to the current point. We update to a proposed point if it returns a permutation with a better value for our objective function than the current point. Note we could also randomise the updating process by incorporating acceptance probabilities that are high if the proposal point is better than the current point and vice versa.

Before giving the full sampling strategy we point out that not all steps may be required, these variants are described in the notes that follow.
\subsection{The Algorithm}

\begin{enumerate}
  \item Initialise \texttt{totalIterations}. Choose $\lambda, M, L$ (see equations (\ref{eq:Qdash}),  (\ref{eqn.Pmax}),  (\ref{eq:ypresamps})), $T,$ target function $f_t$ and acceptance probability function ${\mathcal{A}}(E, E_\star, t)$ (see note~\ref{note:one} below). Denote the dimension of the problem by $n$. 

\medskip Preliminary calculations:
\medskip
  \item Find a relaxed $Q$ minimising (\ref{eqn.GMPrelax}) (if $Q$ is to be doubly stochastic, this can be done using the Frank-Wolfe algorithm).
  \item If $Q\bfa=c \bfa$ for $c \in \mathbb{R}$, where $\bfa = n^{-1/2}\bf1$ (i.e. if $Q$ is doubly stochastic), then perturb $Q$ so that for any permutation $P$, $S(P)$ has no zero measure. To do this, set $Q \leftarrow Q + \lambda U$ \label{step.Perturb}.
  \item Find initial point $\bfx_0$ by reversing the permutation $P_0 = \arg \min_{P \in \mathbb{P}} ||Q-P||^2_{\rm F},$ solved by the Hungarian algorithm. See the discussion around (\ref{eq:plugin}). \label{step.Initialise}
  \item Calculate the associated $E_0 {\displaystyle{\EqualDef}} || A - P_0^T B P_0 ||_{\rm F}$ to give the initial vector $(\bfx_0, P_0, E_0)$.
  \item Randomly draw 
$\bfz_1,\ldots,\bfz_M$ on the unit hpersphere,  and 
calculate ${\Delta}_{\rm max}$ via (\ref{eqn.Pmax}). 
  \item As in section~\ref{subsec:presamples} we generate presamples $y_{-1},\ldots, y_{-L}$ and calculate the associated $\Pdiff_{-1},\ldots,\Pdiff_{-L}$ using (\ref{eq:minusis}) and hence learn $\tmodelP(\sigma^2_t)$; rescale to give $\modelP(\sigma^2_t).$ 

\medskip Main iterations:
\medskip

  \item\label{item:no} For $t = 1: \tt{totalIterations}$
  \item If $t \mod T = 0$, re-learn $\tmodelP(\sigma^2_t)$ based on the new observations; rescale to give $\modelP(\sigma^2_t).$ 
  \item Choose $\sigma^2_t$ minimising $||\modelP(\sigma^2_t)- f_t||^2_{\rm F}$.
  \item Sample $\bfx_\star {\displaystyle{\EqualDist}} {\cal N}(\bfx_{t-1}, \sigma^2_t)$ and normalise $\bfx_\star \leftarrow \frac{\bfx_\star}{||\bfx_\star||}$.
  \item Find corresponding permutation representing a `rounding' of $P,$ namely $P_\star = \arg \min_{P \in \Pset} ||Q \bfx_\star - P \bfx_\star||^2_{\rm F},$ and also $E_\star = ||A - P_\star^T B P_\star||_{\rm F}$.
  \item Sample $u \sim \text{Unif}[0,1]$ and if $u \leq \mathcal{A}(E_{t-1}, E_\star, t)$, set ($\bfx_t$, $P_t$, $E_t$) to ($\bfx_\star$, $P_\star$, $E_\star$) otherwise to ($\bfx_{t-1}$, $P_{t-1}$, $E_{t-1}$).
  \item Loop back to (\ref{item:no}).
\end{enumerate}

\newcounter{note}
\begin{list}
{Note~\arabic{note}:}{\usecounter{note}
}
\item \label{note:one}
The pure strategy choice of ${\mathcal A}$ would be ${\mathcal A}(E, E_\star, t) = {\cal I}(E_\star \leq E)$ where ${\cal I}$ is the indicator function. The effect of  `$\leq$' is to avoid getting stuck in the middle of a large permutation set $S(P)$ as the algorithm will continue to move around inside it.
\item
The choice of the target function $f_t$ is important. If it
decreases too sharply there will not be a chance to sufficiently explore the space of permutation sets; too slowly and there will not be time  to make the small adjustments necessary to update a good permutation to a better one, (based on the  continuity argument), and it becomes too much like random sampling.
\item
The algorithm is suitable for parallelization.  
Consider the {\tt{totalIterations}} to be one run. Different exploring threads could be initiated within one run of the algorithm that re-align at regular intervals based on the current best thread.
\item In step \ref{step.Initialise}, we could also simply use a random value
for $\bfx_0$  in $\mathbb{R}^n$ as our initialisation but it does not normally perform as well.
\end{list}

\section{Results} \label{sec.Results}
Here we use our sampling strategy, rather than  the standard projection step given by (\ref{eqn.matrixRoundCurrent}), with the simple method of finding the optimal doubly stochastic matrix $Q$ solving (\ref{eqn.GMPrelax}). We call our overall method SSQCV. 

In the results which follow  we use $\lambda$ to perturb $Q$ such that $Q \leftarrow Q + \lambda U$ where $U$ is a matrix of uniform random numbers between $[0,1]$. We take
${\tt totalIterations} = 100000; M = 100, L = 1000, \lambda=0.1, T = {\tt totalIterations}/10.$ We use the pure strategy choice of ${\mathcal A}$
and set $f_t=\Pdiff_{\rm max} [1-(t/{\tt totalIterations})^{0.6}].$

Our sampling strategy is applied to the QAPLIB benchmark library used also by \cite{Schellewald_etal01,Zaslavskiy_etal09}.
Now 
\begin{eqnarray*}
&&\!\!\!\!\!\!\!\!\!\!\!\!\!\!\!\!\min_P || A - P^TBP ||_{\rm F}^2 = \min_P \tr[(A - P^TBP)^T(A - P^TBP)]\\
 &=& \min_P \tr[ A^T A - 2A^TP^TBP + P^TB^TBP] \\
&=& \max_P \tr[A^TP^TBP] ,
\end{eqnarray*}
since $P^TB^TBP$ is a permutation of both the rows and columns of $B^TB$ so all elements on the leading diagonal remain on the leading diagonal i.e., its trace is independent of $P$.

As pointed out in \cite{Schellewald_etal01} the quantity  $\tr[A^TP^TBP]$ is negative for this class of experiments so that $\max_P \tr[A^TP^TBP]=
\min_P \tr[-A^TP^TBP]$ where $\tr[-A^TP^TBP]$ values are positive. It is these latter positive values which are displayed in 
Table~\ref{tab:QAPLIB}.

In Table~\ref{tab:QAPLIB} we have
\begin{enumerate}
  \item QAP: The name of the benchmark in QAPLIB.
  \item Min: The true minimum trace value of the benchmark.
  \item PATH: The minimum trace value found by the PATH algorithm.
  \item SSQCV Mean: Over 20 runs of the algorithm, the mean minimum trace value found.
  \item SSQCV Best: Over 20 runs of the algorithm, the best minimum trace value found.
  \item SSQCV Time: Over 20 runs of the algorithm, the mean execution time taken.
\end{enumerate}
\begin{table*}[t!]
\begin{center}
\begin{tabular}{|c|c|c|c|c|c|}
\hline 
QAP & Min & PATH & SSQCV Mean & SSQCV Best & SSQCV Time (s) \tabularnewline
\hline  
\hline 
chr12c & 11156 & 18048 & {\bf 13088}  & 11414  & 15.98 \tabularnewline
\hline 
chr15a & 9896 & 19086 & {\bf 14247} & 11168 & 20.07  \tabularnewline
\hline 
chr15c & 9504 & 16206 & {\bf 15199}  & 11200  & 19.07  \tabularnewline
\hline 
chr20b & 2298 & 5560 & {\bf 3960} & 3054  & 16.73 \tabularnewline
\hline 
chr22b & 6194 & 8500 & {\bf 7574} & 7196  & 17.50 \tabularnewline
\hline 
exc16b & 292 & 300 & {\bf 292} & 292 & 16.54 \tabularnewline
\hline 
rou12 & 235528 & 256320 & {\bf 246063} & 240598 & 16.31 \tabularnewline
\hline 
rou15 & 354210 & 391270 & {\bf 380746}  & 365264  & 16.49 \tabularnewline
\hline 
rou20 & 725522 & {\bf 778284} & 778709 & 760874  & 16.99 \tabularnewline
\hline 
tai15a & 388214 & 419224 & {\bf 409769}  & 395714  & 16.94 \tabularnewline
\hline 
tai17a & 491812 & 530978 & {\bf 525815}  & 514496  & 16.76 \tabularnewline
\hline 
tai20a & 703482 & {\bf 753712} & 766274  & 751414  & 17.03 \tabularnewline
\hline 
tai30a & 1818146 & {\bf 1903872} & 1979579  & 1946888  & 18.37 \tabularnewline
\hline 
tai35a & 2422002 & {\bf 2555110} & 2659594  & 2613758  & 22.40 \tabularnewline
\hline 
tai40a & 3139370 & {\bf 3281830} & 3459139 & 3407476 & 24.16 
\tabularnewline\hline
\end{tabular}
\end{center}
\caption{Experimental results for QAPLIB benchmark data sets}
\label{tab:QAPLIB}
\end{table*}

We make the following observations:
\begin{itemize}
\item
SSQCV Mean  outperforms PATH in two-thirds of the experiments. We already know from \cite{Zaslavskiy_etal09} that PATH outperforms competitors such as QPB \cite{Schellewald_etal01}, GRAD \cite{GoldRangarajan96}, or Umeyama's algorithm \cite{Umeyama88}.
\item
The PATH algorithm tends to perform better at higher dimensions. This is due to the fact $\mathbb{R}^n$ becomes more and more finely partitioned as dimension increases and we need more iterations and a slower decrease in variance in our algorithm to account for this. This is the point where the benefits of updating of $Q$ in PATH begins to outweigh the benefits of the sampling strategy with a fixed $Q$.
\item
While this is a comparison against the PATH algorithm, we note that the sampling strategy can be integrated with more complex methods to achieve better results, including the PATH algorithm itself. As the dimension increases, it becomes clear that the partitioned space generated by $Q$ does not have enough  `large' sets $S_Q(P)$ where $P$ is a good solution to (\ref{eqn.GMPrelax}). This suggests that an approach that also iteratively updates $Q$ (as in PATH) would produce better results with our sampling strategy. Of course, using the sampling strategy with the PATH algorithm would provide the best of both worlds in terms of performance.
\end{itemize}
\appendix
\subsection{Proof of Theorem~\ref{thm.minSort}}
\label{proof:ProofminSort}
	We first show that 
	$$ F(P) \EqualDef ||\bfa - P \bfb||^2_F $$
	is minimised when permutation matrix $P$ sorts the vector $\bfb$ such that $a_i \leq a_j \Rightarrow (P \bfb)_i \leq (P \bfb)_j$ i.e., $\bfr(\bfa) = \bfr(P \bfb)$.

The contribution to $F$ at indices $i$ and $j$ is
	$$ (a_i - c_i)^2 + (a_j - c_j)^2$$
where $\bfc \,{\displaystyle{\EqualDef}}\, P \bfb.$
Now,
\begin{eqnarray*}
	(a_i - c_i)^2 &=& (a_i - c_j + c_j - c_i)^2\\
	 &=& (a_i - c_j)^2 + (c_j - c_i)(2a_i - c_j - c_i).
\end{eqnarray*}	
Similarly,
	$$ (a_j - c_j)^2 = (a_j - c_i)^2 + (c_i - c_j)(2a_j - c_j - c_i).$$	
Therefore,
	\begin{eqnarray}\label{eqn.sortProp1}
		(a_i - c_i)^2 + (a_j - c_j)^2 &=& (a_i - c_j)^2 + (a_j - c_i)^2\nonumber\\
&+& (c_j - c_i)(2a_i - 2a_j).
\end{eqnarray}

We also know that if $P$ is to be an optimal transformation, we must have
\begin{equation}\label{eqn.sortProp2}
		(a_i - c_i)^2 + (a_j - c_j)^2 \leq (a_i - c_j)^2 + (a_j - c_i)^2,
\end{equation}
otherwise we can define $P^{'}$ such that $(P^{'} \bfb )_k = (P \bfb )_k$ for $k \neq i,j$ but $(P^{'} \bfb )_i = (P \bfb )_j$ and $(P^{'} \bfb )_j = (P \bfb )_i$. Clearly if (\ref{eqn.sortProp2}) did not hold, $F(P^{'}) < F(P)$, contradictory to $P$ being optimal.

Combining (\ref{eqn.sortProp1}) and (\ref{eqn.sortProp2}) gives
$ (c_j - c_i)(a_i - a_j) \leq 0.$
Hence if $a_i \leq a_j$ then we must have $c_i \leq c_j$. Thus $F(P)$ is minimised when $P$ sorts $\bfb$ to the same ordering as $\bfa$.
Letting $\bfa = Q \bfx$ and $\bfb = \bfx$ in Theorem~\ref{thm.minSort} gives the result.

\subsection{Proof of Proposition~\ref{prop:meanequiv}}
\label{proof:ProofExpectation}
We show this in 3 steps:
	\begin{enumerate} 
		\item Firstly,
$$\min_{P \in \Pset} E( || Q \bfX - P \bfX ||^2_{\rm F}) = \max_{P \in \Pset} \int_{\UnitH} \bfx^T Q^T P \bfx\, \dif \bfx.$$
		\item Considering the quantity $\int_{\UnitH} \bfx^T A \bfx \,\dif\bfx$ for some $A \in {\mathbb{R}}^{n \times n},$
		\begin{enumerate}
			\item all off-diagonal terms, i.e., those of the form $A_{ij} x_i x_j$ for $i \neq j,$ integrate to 0,
			\item all diagonal elements $A_{ii} x_i^2$ integrate to $A_{ii} \beta$ for some constant $\beta$.
		\end{enumerate}
		\item Hence $\max_{P \in \Pset} \int_{\UnitH} \bfx^T Q^T P \bfx\, \dif \bfx$ is equivalent to maximising $\tr(Q^TP)$ which is equivalent to $\min_{P \in \Pset} ||Q-P||_{\rm F}$, so the result in (\ref{eqn.unifSphere}) follows.
	\end{enumerate}
	
\subsubsection{Step 1}
\begin{eqnarray*}
	||Q \bfx - P \bfx||^2_{\rm F} &=& \tr \{(Q \bfx - P \bfx)^T (Q \bfx - P \bfx)\}\\
	&=& \tr\{(Q \bfx)^T (Q \bfx)\} + \tr\{(P \bfx)^T (P \bfx)\}\\
 &-& 2\bfx^T Q^T P \bfx \\
	&=& 2 \bfx^T \bfx - 2 \bfx^T Q^T P \bfx,
\end{eqnarray*}
using the fact that both $Q$ and $P$ are orthogonal matrices.

Therefore,
\begin{eqnarray*}
	\min_{P \in \Pset} E( || Q \bfX - P \bfX ||^2_{\rm F}) &=& \max_{P \in \Pset} \int_{\UnitH} \bfx^T Q^T P \bfx \,\dif \mu\\
 &=& \max_{P \in \Pset} \int_{\UnitH} \bfx^T Q^T P \bfx \,\dif \bfx,
\end{eqnarray*}
where the final equality is a result of $\mu$ being a uniform distribution.
	
\subsubsection{Step 2}
	Now consider $I\,\, {\displaystyle{\EqualDef}} \,\,\int_{\UnitH} \bfx^T A \bfx \dif \bfx.$
Writing $\bfx$ in terms of hyperspherical coordinates, we have on the unit hypersphere that the volume element is
$$\sin^{n-2} (\theta_1) \sin^{n-3} (\theta_2) \ldots \sin (\theta_{n-2}) \dif \theta_1 \dots \dif \theta_{n-1},
$$
and
\begin{eqnarray*}
	x_1 &=& \cos(\theta_1)\\
	x_2 &=& \sin(\theta_1) \cos(\theta_2)\\
	&\vdots& \\
	x_{n-1} &=& \sin(\theta_1) \ldots \sin(\theta_{n-2}) \cos(\theta_{n-1})\\
	x_{n} &=& \sin(\theta_1) \ldots \sin(\theta_{n-2}) \sin(\theta_{n-1}).
\end{eqnarray*}	
	Consider off-diagonal elements of $I$ of the form
$A_{ij} x_i x_j.$ For $i \neq j$, we see that $x_i x_j$ contains at least one term of the form $\sin^L (\theta_k) \cos(\theta_k)$ for $L \geq 0,$ i.e., when $k=i$ or $k=j$ as we cannot have both $i=n$ and $j=n$ as they cannot be equal. Hence
	$$ \int_{\mathbb{X}} x_i x_j \,\dif \bfx = 2 \int_{\mathcal{I}} \int_{0}^{\pi} \sin^L (\theta_k) \cos(\theta_k) \dif \theta_k f(\bftheta_{/k}) \dif \bftheta_{/k} $$
	where $f$ is some function, $\bftheta_{/k}$ is a vector of all $\theta_l$ without $\theta_k$ and $\mathcal{I}$ is the region over which we are integrating $\bftheta_{/k}.$ 
But,
	$$\int_{0}^{\pi} \sin^L (\theta_k) \cos(\theta_k) \dif \theta_k = \left [ \frac{\sin^{L+1}(\theta_k)}{L+1} \right ]_0^{\pi} = 0,
$$
so all off-diagonal elements of $I$ integrate to 0.
	
Now consider diagonal elements of $I,$ of the form
$A_{ii} x_i^2.$
We now require two identities. Firstly,
	\begin{equation} \label{eqn.Identity1}
		\int_0^\pi \sin^n(\theta) \cos^2(\theta) \dif \theta = \int_0^\pi \frac{\sin^{n+2}(\theta)}{n+1} \dif \theta,
	\end{equation}
	found from integrating by parts with $\dif v = \sin^n(\theta)\cos(\theta)$ and $u = \cos(\theta)$. Secondly,	
	\begin{equation} \label{eqn.Identity2}
		\int_0^\pi \sin^n(\theta) \dif \theta = \frac{n-1}{n} \int_0^\pi \sin^{n-2}(\theta) \dif \theta
	\end{equation}
	integrating by parts with $\dif v = \sin(\theta)$ and $u = \sin^{n-1}(\theta)$.
	
	For $i \neq n$, we see that $\int_{\UnitH} x_i^2 \dif \bfx$ can be written
	\begin{eqnarray} 
\!\!\!\!\int_{\UnitH} x_i^2 \dif \bfx\!\!\!\!\! &=&\!\!\!\!\!	2 \int_0^\pi [\sin^2(\theta_1) \ldots \sin^2(\theta_{i-1}) \cos^2(\theta_i)]\nonumber\\
 \!\!\!\!\!\!\!\!\!\!&\times&\!\!\!\!
 \sin^{n-2}(\theta_1) \sin^{n-3}(\theta_2) \ldots \sin(\theta_{n-2}) \dif \bftheta,
		\label{eqn.intSquared}
	\end{eqnarray}	 
where $\int_0^\pi$ represents the fact all $\theta_l$ are to be integrated between these bounds.
	
	Now consider $\theta$-index $j$ and define $\int_0^\pi \sin^k(\theta) \dif \theta {\,\,\displaystyle{\EqualDef}\,\,}  I_k$.
	
Case 1: $j<i$
The relevant integral in (\ref{eqn.intSquared}) is
$$ \int_0^\pi \sin^{n-j-1}(\theta_j) \sin^2(\theta_j) \,\dif\theta_j = I_{n-j+1} = \frac{n-j}{n-j+1} I_{n-j-1}, 
$$
using (\ref{eqn.Identity2}).
	
Case 2: $j>i$
The relevant integral in (\ref{eqn.intSquared}) is
$$ \int_0^\pi \sin^{n-j-1}(\theta_j) \dif \theta_j = I_{n-j-1}.
 $$
	
	Case 3: $j=i$
	
	The relevant integral in (\ref{eqn.intSquared}) is
\begin{eqnarray*} \int_0^\pi \cos^2(\theta_j) \sin^{n-j-1}(\theta_j) \,\dif\theta_j &=& \frac{1}{n-j} I_{n-j+1} \\
&=& \frac{1}{n-j+1} I_{n-j-1}. 
\end{eqnarray*}
using both (\ref{eqn.Identity1}) and (\ref{eqn.Identity2}).
	
Putting together all these cases we see that 
\begin{eqnarray*} \int_{\UnitH} \!\!\!\!x_i^2 \,\dif \bfx \!\!\!\!&=& \!\!\!\!
2 \left( \prod_{j=1}^{i-1} \frac{n-j}{n-j+1} I_{n-j-1} \right) \left( \prod_{j=i+1}^{n-1} I_{n-j-1} \right)\\
&\times& \left( \frac{1}{n-i+1} I_{n-i-1} \right) \\
	 &=& 2 \left( \frac{n-1}{n} \frac{n-2}{n-1} \cdots \frac{n-i+1}{n-i+2}\frac{1}{n-i+1} \right) \\
&\times&\prod_{j=1}^{n-1} I_{n-j-1} \\
	 &=& \frac{2}{n} \alpha, 
\end{eqnarray*}
	where $ \alpha = \prod_{j=1}^{n-1} I_{n-j-1} = \prod_{j=0}^{n-2} I_{j}. $ 
	
	Finally we look at $i=n$, for which $\int_{\UnitH} x_n^2 \,\dif\bfx$ is
\begin{eqnarray*}
 \!\!\!2 \int_0^\pi &&\!\!\!\!\!\!\!\!\!\!\!\!\!\!\!\sin^n(\theta_1) \sin^{n-1}(\theta_{2}) \ldots \sin^{3}(\theta_{n-2}) \sin^{2}(\theta_{n-1}) \,\dif \bftheta \\
	 \!\!\!\!\!&=&\!\!\!\!\! 2 \prod_{j=2}^n I_j 
	 = 2 \prod_{j=2}^n \frac{j-1}{j} I_{j-2} 
	 = \frac{2}{n} \prod_{j=0}^{n-2} I_j 
	 = \frac{2}{n} \alpha.
\end{eqnarray*}
	Hence,
	$$ \int_{\UnitH} \bfx^T A \bfx \,\dif \bfx = \int_{\UnitH} \sum_{i=1}^n A_{ii} x_i^2 \,\dif \bfx = \frac{2}{n} \alpha \tr\{A\}. 
$$
\subsubsection{Step 3}
Now we see that,
\begin{eqnarray*}\!\!\!\! \arg \min_{P \in \Pset} E(||Q \bfX - P \bfX||^2_{\rm F}) \!\!\!\!\!&=&\!\!\!\!\! \arg \max_{P \in \Pset} \int_{\mathbb{X}} \bfx^T Q^T P \bfx \,\dif \bfx \\
 \!\!\!\!\!&=&\!\!\!\!\! \arg \max_{P \in \Pset} \tr\{Q^T P\}\\
 \!\!\!\!\!&=&\!\!\!\!\! \arg \min_{P \in \Pset} ||Q-P||^2_{\rm F},
\end{eqnarray*}
when $\bfX$ is uniformly distributed on the unit hypersphere.

\subsection{Proof of Proposition~\ref{prop:hypercube}}
\label{proof:ProofHypercube}
		In this case we have for $A = Q^T P$,
$$ \int_H \bfx^T A \bfx \,\dif \bfx = \int_H \left( \sum_{i=1}^n A_{ii} x_i^2 + \sum_{i \neq j} A_{ij} x_i x_j \right) \,\dif \bfx $$
	$$ = \left[ \frac{1}{3} \sum_{i=1}^n A_{ii} x_i^3 \frac{V}{x_i} + \frac{1}{4} \sum_{i \neq j} A_{ij} x_i^2 x_j^2 \frac{V}{x_i x_j} \right]_H$$
where $V = \prod_{i=1}^n x_i$.
Plugging in the limits for $H$ the inegral is 
$$ \frac{1}{3} \tr\{A\} + \frac{1}{4} {\bf 1}^T A {\bf 1} - \frac{1}{4} \tr\{A\} = \frac{1}{12} \tr\{A\} + \frac{1}{4} {\bf 1}^T A {\bf 1}. $$
	
	Noting that ${\bf 1}^T Q^T P {\bf 1} = {\bf 1}^T Q^T {\bf 1}$ is invariant for all permutation matrices as they simply permute the columns of $Q^T$, we see that	
	$$ \arg \min_{P \in \Pset} E(||Q \bfX - P \bfX ||^2_{\rm F}) = \arg \max_{P \in \Pset} \tr\{Q^T P\} $$
	and the result follows.

\subsection{Proof of Proposition~\ref{prop.closedSet}}\label{app:closedSet}
Using Theorem~\ref{thm.minSort}, we know that for the permutation to be the same for $\bfx$ and $\bfy$
it is sufficient, (from (\ref{eqn.Barvinok})), that 
$$ \bfr(\bfx) = \bfr(\bfy)\,\,\text{and}\,\,\bfr(Q \bfx)=\bfr(Q \bfy).
$$
We have $y \in {\mathbb{B}}_{\epsilon}(\bfx)$ such that by definition $||\bfy - \bfx||^2_{\rm F} < \epsilon.$ Also, $|| Q (\bfy - \bfx)||^2_{\rm F} < \delta \epsilon$, where $\delta=\max\{ |\lambda|:\lambda\text{ is an eigenvalue of }Q^TQ\},$ \cite[p.~296]{HornJohnson85}.

By definition no $(Q \bfx)_i = (Q \bfx)_j,$ so for the sorting order to remain the same for $Q\bfy$ and $Q \bfx$ we require that, if
$ (Q \bfx)_i < (Q \bfx)_j, $
then 
$ (Q \bfy)_i <  (Q \bfy)_j.$
	
Also note that if, $x_i < x_j$
then for the sorting to remain the same, we require
$y_i < y_j.$
	
Now, $|y_i-x_i|< \epsilon^{1/2}$ so $y_i\in (x_i-\epsilon^{1/2},x_i+\epsilon^{1/2}) ,$ so
	$$ y_i < x_i + \epsilon^{1/2} $$	
and, similarly,
	$$ y_j > x_j - \epsilon^{1/2}. $$	
Then taking for example
	$$ \epsilon^{1/2}=\epsilon_1 < \frac{1}{2} \min_{u,v} | x_u - x_v | $$
ensures that	
	$$ y_i < x_i + \epsilon_1 < x_j - \epsilon_1 <  y_j. $$
Similarly
	$$ (Q \bfy)_i < (Q \bfx)_i + (\delta \epsilon)^{1/2} $$
and
	$$ (Q \bfy)_j > (Q \bfx)_j - (\delta \epsilon)^{1/2}. $$
Then taking 
$$ \epsilon_2=\epsilon^{1/2} < \frac{1}{2\delta^{1/2}} \min_{u,v} | (Q \bfx)_u - (Q \bfx)_v | $$
ensures that	
\begin{eqnarray*} (Q \bfy)_i &<& (Q \bfx)_i + \delta^{1/2} \epsilon_2 < (Q \bfx)_j - \delta^{1/2} \epsilon_2\\
 &<& (Q \bfy)_j.
\end{eqnarray*}
Hence we can choose
	$ \epsilon = \min (\epsilon_1^2, \epsilon_2^2),$ completing the proof.

\subsection{Proof of Theorem~\ref{thm.reversePerm}}\label{app:reversePerm}
Let $\bfx = \bfb + A \bfepsilon$ for some $A \in {\mathbb{R}}^{n \times n}$. Now, $Q \bfx = Q (\bfb + A \bfepsilon) = \bfa + Q A \bfepsilon$.

Step 1. What is $\bfr (\bfx)$?
Since $b_i = b_j \Rightarrow i = j$, we can choose $\delta > 0$ sufficiently small such that
\begin{equation}\label{eq:rxrb} \bfr (\bfx) = \bfr (\bfb),
\end{equation}
i.e., the ordering of $\bfx$ is the same as $\bfb$. Note that $\bfr (\bfepsilon) = \bfr (P_b \bfb)$ as both are in ascending order.

Step 2. What is $\bfr ( Q \bfx )$?
Now $\bfr (Q \bfx) = \bfr ( \bfa + Q A \bfepsilon ) = \bfr ( QA \bfepsilon ) $ as $ \bfa $ is a constant vector. We can therefore choose $A = Q^{-1} P^* P_b^T$ to get
$$ \bfr(Q \bfx) = \bfr ( P^* P_b^T \bfepsilon) = \bfr (P^* \bfb ). $$
Then we use (\ref{eq:roundstoself}) which says that $\bfr (P^* \bfb )=P^*\bfr (\bfb ).$ So
$$
\bfr(Q \bfx) = P^*\bfr (\bfb )=P^*\bfr(\bfx),
$$
where the last step uses (\ref{eq:rxrb}).

Step 3. Therefore by Theorem~\ref{thm.minSort}, for $\bfx = \bfb + Q^{-1} P^* P_b^T \bfepsilon$, $P^* = \arg \min_{P \in \mathbb{P}} |Q \bfx - P \bfx|$, and the proof is complete.

\nocite{*}
\bibliographystyle{IEEE}

\end{document}